\documentclass{elsart5p}
\usepackage{graphicx,natbib,amssymb}
\journal{New Astronomy}
\newcommand{\kpc}{\,{\rm kpc}}
\newcommand{\kms}{\,km\,s$^{-1}$}
\newcommand{\cmt}{\,cm$^{-3}$}
\newcommand{\cmd}{\,cm$^{-2}$}     
\newcommand{\myr}{\,$M_{\odot}\,{\rm yr}^{-1}$}
\newcommand{\es}{$\,\rm erg\,s^{-1}$}
\newcommand{\ecs}{$\,\rm erg\,cm^{-2}\,s^{-1}$}
\newcommand{\ecsa}{$\,\rm erg\,cm^{-2}\,s^{-1}\,\AA^{-1}$}
\newcommand{\ha}{H$\alpha$}

\newcommand{\oi}{O\,{\small I}}
\newcommand{\ro}{\,$R_{\odot}$}
\newcommand{\mo}{\,$M_{\odot}$}
\newcommand{\lo}{\,$L_{\odot}$}
\begin{document}
\begin{frontmatter}

\vspace*{-4cm}

\title{Multiwavelength modelling the SED of supersoft X-ray sources \\[1mm]
       III. RS~Ophiuchi: The supersoft X-ray phase and beyond}

\author{A.~Skopal\thanksref{fn1}, \thanksref{fn2}}

\thanks[fn1]{E-mail: skopal@ta3.sk}
\thanks[fn2]{Visiting Astronomer: Astronomical Institut, Bamberg}
\thanks[fn3]{http://dx.doi.org/10.1016/j.newast.2014.05.008}

\address{Astronomical Institute, Slovak Academy of Sciences,
         059\,60 Tatransk\'{a} Lomnica, Slovakia \\[1mm]
        {\rm Received 17 July 2013; accepted 20 May 2014}
        }
\vspace*{5mm}

{\small
\hspace*{-15cm}
{\rm \bf H I G H L I G H T S}\\
\begin{itemize}
\item
\mbox{Multiwavelength model SEDs of the nova RS~Oph from 
      X-rays to mid-IR were performed. }
\item
\mbox{During the supersoft source phase its luminosity was 
      highly super-Eddington. }
\item
\mbox{It was sustained by a high accretion rate from a disk, 
      followed by jets. }
\item
\mbox{During quiescence the SED satisfied radiation produced 
      by a large accretion disk. }
\item
\mbox{The high accretion rate could be realized throughout 
      a focused wind of the giant. }
\end{itemize}
}
\vspace*{-7mm}
 
\begin{abstract}
I modelled the 14\,\AA --37\,$\mu$m SED of the recurrent 
symbiotic nova RS~Oph during its supersoft source (SSS) phase 
and the following quiescent phase. 
During the SSS phase, the model SEDs revealed the presence of 
a strong stellar and nebular component of radiation in the 
spectrum. The former was emitted by the burning WD at highly 
super-Eddington rate, while the latter represented a fraction 
of its radiation reprocessed by the thermal nebula. 
During the transition phase, both the components were decreasing 
and during quiescence the SED satisfied radiation produced by 
a large, optically thick disk ($R_{\rm disk} > 10$\ro). 
The super-Eddington luminosity of the burning WD during the SSS 
phase was independently justified by the high quantity of the 
nebular emission. 
The emitting material surrounded the burning WD, and its mass 
was $(1.6\pm 0.5)\times 10^{-4}(d/1.6\kpc)^{5/2}$\mo. 
The helium ash, deposited on the WD surface during the whole
burning period, was around of 
  $8\times 10^{-6}(d/1.6\kpc)^2$\mo, 
which yields an average growing rate of the WD mass, 
  $\dot M_{\rm WD} \sim 4\times 10^{-7}(d/1.6\kpc)^2$\myr. 
The mass accreted by the WD between outbursts, 
  $m_{\rm acc} \sim 1.26\times 10^{-5}$\mo, 
constrains the average accretion rate, 
  $\dot M_{\rm acc} \sim 6.3\times 10^{-7}$\myr. 
During quiescence, the accretion rate from the model SED of 
$\sim 2.3\times 10^{-7}$\myr\ requires a super-Eddington 
accretion from the disk at $\sim 3.6\times 10^{-5}$\myr\ during 
the outburst. Such a high accretion can be responsible for the 
super-Eddington luminosity during the whole burning phase. 
Simultaneous presence of jets supports this scenario. 
If the wind from the giant is not sufficient to feed the WD 
at the required rate, the accretion can be realized from the 
disk-like reservoir of material around the WD. In this case 
the time between outbursts will extend, with the next explosion 
beyond 2027. 
In the opposite case, the wind from the giant has to be focused 
to the orbital plane to sustain the high accretion rate at 
a few $\times 10^{-7}$\myr. Then the next explosion 
can occur even prior to 2027. 
%
%
\end{abstract}
\end{frontmatter}
%
%
\section{Introduction}

The last outburst of the recurrent symbiotic nova RS~Oph was 
discovered by \cite{narumi+06} on 2006 February 12.83 UT. 
Due to its expectation and a bright peak magnitude of 
$V = 4.5$, a large amount of observations at different dates 
of the nova evolution, throughout a very wide wavelength 
range was carried out by telescopes from the ground and 
onboard the satellites 
\citep[e.g.][and references therein]{evans07}. 
During the first 30 days after the eruption, observations 
indicated a non-spherical shaping of the nova ejecta 
\citep[e.g.][in the radio]{tob+06}. 
The multiwavelength modelling the SED from this period revealed 
a biconical ionization structure of the nova that developed 
during the first four days and that the luminosity was well 
above the Eddington limit \citep[][(Paper~II)]{sk15b}. 
Signatures of the blast wave produced in the nova explosion were 
detected by the {\em Swift} and {\em RXTE} satellites in the 
form of a strong 0.5--20\,keV X-rays \citep[][]{bode+06,sok+06}. 
The authors interpreted the behaviour of the hard X-ray flux as 
a result of the evolution of shock systems due to the impact of 
high-velocity ejecta into the red giant wind. 

Around day 26, a luminous and highly variable supersoft X-ray 
component began to appear in the spectrum 
\citep[][]{bode+06,ness+09,osborne+11}. 
RS~Oph thus entered its SSS phase, which lasted to about day 90 
\citep[][]{osb+06}. Between days 40 and 80 the supersoft X-ray 
fluxes were relatively stable \citep[][]{hkl07,osborne+11}. 
This period of the nova evolution provides a good occasion 
to determine the fundamental parameters of the burning WD, 
because of the possibility to measure its radiation throughout 
a broad energy region. 

\cite{ness+07} analyzed first the X-ray \textsl{Chandra} and 
\textsl{XMM-Newton} grating spectra from the SSS phase. 
On day 54, they found the best agreement between their blackbody 
models and observations for the temperature and luminosity 
$T \sim (520-580)\times 10^3$\,K and 
$\log(L) = 39.3-40.0$, respectively, however without 
specifying the corresponding $N_{\rm H}$. They did not claim 
from these models that a super-Eddington luminosity occurred. 

\cite{nelson+08} compared this spectrum with WD atmospheric 
models, and selected that with 
$T \sim 820000 \pm 10000$\,K and 
$N_{\rm H} = 2.3\times 10^{21}$\cmd\ 
for $L \equiv L_{\rm Edd}$. 

\cite{osborne+11} performed spectral fits to the {\em Swift}-XRT 
SSS spectra by the atmosphere models. They obtained 
 $kT \sim 90$\,eV ($\sim 1.04\times 10^{6}$\,K), 
 $R \sim 4\times 10^8$\,cm ($\sim 0.0057$\ro) and 
 $L \sim 3\times 10^4$\lo. 

The very different quantities of fundamental parameters of the 
burning WD during the SSS phase, as derived by different groups 
of authors, result probably from the problem of mutual 
dependence between the parameters $L, N_{\rm H}, T$ in fitting 
only the X-ray data 
\citep[see Sect.~4 of][Paper~I therein]{sk15a}. 
%

In this contribution I aim to determine the fundamental 
($L,R,T$) and $N_{\rm H}$ parameters of the global spectrum 
produced by RS~Oph during the SSS phase and beyond it by 
modelling its SED throughout the very large spectral range, 
0.0014--37\,$\mu$m. 
Section~2 introduces multiwavelength observations at selected 
days, the results of their SED-fitting analysis are given in 
Sect.~3, and discussion with summary are found in Sects.~4 and 5, 
respectively. 

\section{Observations}

As in the Paper~II (see Sect.~2 there) I used observations 
obtained during the two recent, 1985 and 2006, outbursts. 
I complemented the X-ray observations from the SSS phase of 
the 2006 outburst with the nearest (in days after the maximum) 
\textsl{IUE} spectra, the optical $UBVR_CI_C$ and 
the infrared \textsl{JHKL} photometric flux-points from 
the 1985 outburst, and with those obtained by the Infrared 
Spectrometer on the {\em Spitzer Space Telescope} 
\citep[][ DDT, PID\,270]{evans07}. 
Photometric measurements were interpolated to dates of the 
\textsl{IUE} spectra with the aid of light curves (LC) 
published by \cite{evans88}. 

For the SSS phase, I analyzed three X-ray spectra obtained 
with {\em Chandra} (day 40 and 67) and \textsl{XMM-Newton} 
(day 54), i.e. close to the beginning, middle and the end of 
the stable SSS phase. These X-ray spectra were already analyzed 
by \cite{ness+07}, \cite{nelson+08} and \cite{ness+09}, who 
also provided their detailed description and treatment. 
For the aim of this paper I selected around 20 representative 
fluxes from the figure 4 of \cite{ness+07}. 

\begin{figure*}
\centering
\begin{center}
%
\resizebox{\hsize}{!}{\includegraphics[angle=-90]{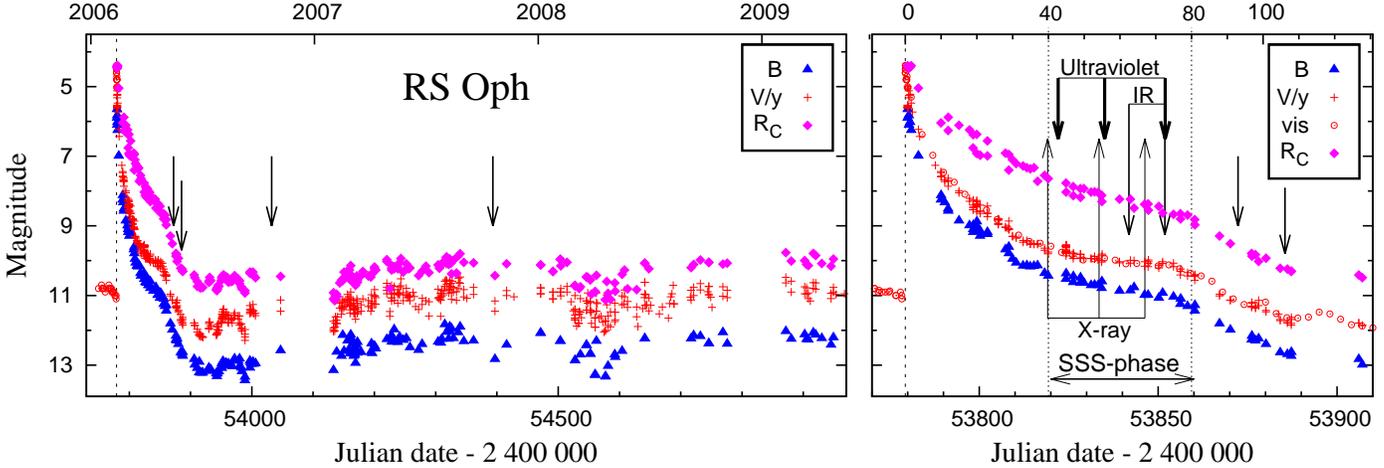}}
\end{center}
\caption[]{
The $BVR_C$ LCs of RS~Oph from the maximum at 2006 February 12.83 
(vertical dashed lines). The data were collected by the VSOLJ 
observers, Kiyota, Kubotera, Maehara and Nakajima. Visual 
estimates are from CDS. The right panel shows a detail covering 
the 2006 outburst. The plateau phase between day 40 and 80 
coincides with a stable SSS phase of the nova evolution. 
Vertical arrows denote timing of the used spectroscopic 
observations (Table~1). 
          }
\label{fig:lcs}
\end{figure*}

Beyond the SSS phase, during the monotonic decline following 
the plateau phase (day$\gtrsim 80$), two \textsl{IUE} spectra 
were exposed on day 93.6 (SWP25815/LWP05865) and 106.1 
(SWP25920/LWP05962). They were complemented with the photometric 
$V$ and $I_{\rm C}$ flux-points. Other photometric fluxes could 
not be used to estimate the true continuum, because of saturation 
by a strong emission line spectrum. 
%
During the post-outburst minimum (Fig.~\ref{fig:lcs}), the 
UV/optical SED was determined from the \textsl{IUE} spectra 
SWP26883/LWP06860 from 06/10/1985 (day 253.3) and the flux 
corresponding to the visual magnitude $12.0\pm 0.5$. 
The short-wavelength part of the LWP spectrum (2000--2400\,\AA) 
was underexposed and thus not used to model the SED (dotted 
line in Fig.~\ref{fig:beyond}). 
%
During the quiescent phase of RS~Oph, which was established 
about 1 year after the outbursts, the SED was determined from 
the \textsl{IUE} spectrum SWP29351/LWP09232 from 02/10/1986 
(day 614.2), $UBVR_CI_C$ and $JHKL$ photometric fluxes from 
30/09/1986 and 26/09/1986, respectively \citep{evans88}. 

Evolution in the optical brightness of the 2006 outburst, 
the period of its stable SSS phase and timing of the used 
spectroscopic observations are shown in Fig.~\ref{fig:lcs}. 
In modelling the data I used the same absorption model, 
colour excess and distance to RS~Oph as in the Paper~II. 
The log of the used observations is given in Table~1. 
%
%
\begin{table}
\caption[]{Log of observations}
\begin{center}
\begin{tabular}{ccccc}
\hline
\hline
~Date~&~~Julian date~~&~~Day$^{a}$~~&~~Region~~&~~Observatory \\
\hline
Mar~15, 2006&2\,453819.08 & 39.76 & 1.40--3.40\,nm &\textsl{Chandra}    \\
Mar~~1, 1985&2\,446134.70 & 42.73 &  115--335\,nm  &\textsl{IUE}        \\
Mar~~1, 1985&2\,446134.70 & 42.73 & $UVI_CJHKL$    &\textsl{SAAO}$^b$   \\
\hline
Mar~29, 2006&2\,453833.49 & 54.16 & 1.50--3.3\,nm  &\textsl{XMM-Newton} \\
Mar~14, 1985&2\,446147.76 & 54.79 &  115--335\,nm  &\textsl{IUE}        \\
Mar~14, 1985&2\,446147.76 & 54.79 & $UVI_CJHKL$    &\textsl{SAAO}$^b$   \\
Apr~~7, 2006&2\,453841.84 & 62.51 &5.28--37.1\,$\mu$m&\textsl{Spitzer}  \\
\hline
Apr~11, 2006&2\,453846.29 & 66.96 & 1.50--3.20\,nm &\textsl{Chandra}    \\
Mar~31, 1985&2\,446164.63 & 72.66 &  115--335\,nm  &\textsl{IUE}        \\
Mar~31, 1985&2\,446164.63 & 72.66 & $UVI_CJHKL$    &\textsl{SAAO}$^b$   \\
Apr~17, 2006&2\,453851.87 & 72.54 &5.28--37.1\,$\mu$m&\textsl{Spitzer}  \\
\hline
Apr~21, 1985&2\,446185.61 & 93.64 & 115--335\,nm   &\textsl{IUE}        \\
May~~3, 1985&2\,446198.07 &106.1  & 115--335\,nm   &\textsl{IUE}        \\
Sep~27, 1985&2\,446345.25 &253.3  & 115--335\,nm   &\textsl{IUE}        \\
Sep~23, 1986&2\,446706.14 &614.2  & 115--335\,nm   &\textsl{IUE}        \\
\hline
\hline
\end{tabular}
\end{center}
\vspace*{2mm}
$^{a}$~=\,JD -- JD$_{\rm max}$ 
         (JD$_{\rm max}$~2\,446\,091.97 as on 1985 Jan. 26.47; \\
\hspace*{22mm}
          JD$_{\rm max}$~2\,453\,779.33 as on 2006 Feb. 12.83) \\
$^{b}$~values from \cite{evans88} interpolated to JD(\textsl{IUE})
\end{table}
%
%
%
%
%
%
\begin{table*}
\begin{center}
\scriptsize 
\caption{Parameters of the X-ray---IR SED-fitting analysis 
         (Sect.~3, Figs.~\ref{fig:seds} and \ref{fig:beyond}) 
        }
\begin{tabular}{ccccccccccc}
\hline
\hline
~Day~                      &
~$\xi$~                    &
~$N_{\rm H}$~              &
~$T_{\rm h}$~              &
~$\theta_{\rm h}/10^{-12}$~&
~~~$R_{\rm h}^{\rm eff}$~~~    &
~~~$\log(L_{\rm h}$)~~~        &
~~~$L_{\rm ph}({\rm H})$~~~    &
~~$T_{\rm e}$~~              &
~~~~$EM$~~~~                     &
~~$\chi^2_{\rm red}$ / d.o.f.  \\
                     &
                     &
[$10^{21}$\cmd]      &
[kK]                 &
                     &
[$R_{\odot}$]         &
[\es]                &
[s$^{-1}$]           &
[K]                  &
[cm$^{-3}$]          &
                     \\
%
%
\hline
40$^{\star}$
   &$\sim$0.25& 7.3$\pm 0.2$ &$460\pm 10$ 
   & 3.9$\pm 0.7$ & 0.28$\pm 0.04$ 
   & 40.08$\pm 0.13$ &6.8$\times 10^{49}$&36,000 
   & 3.4$\times 10^{61}$&5.9 / 38 \\
54$^{\star}$
   &$\sim$0.6& 6.8$\pm 0.3$ &$495\pm 20$
   & 3.0$\pm 0.6$ & 0.21$\pm 0.05$
   & 39.98$\pm 0.18$ &5.0$\times 10^{49}$&29,000 
   & 2.2$\times 10^{61}$& 4.9 / 42 \\
67$^{\star}$
   &$\sim$0.4& 7.0$\pm 0.2$ &$505\pm 10$
   & 2.7$\pm 0.5$ & 0.19$\pm 0.03$
   & 39.91$\pm 0.11$ &4.3$\times 10^{49}$&26,000 
   & 9.8$\times 10^{60}$&5.5 / 38 \\
\hline
93$^{\dagger}$ 
   &              &          &$> 90^{a}$
   & $< 4.0$      & $< 0.29$ & $> 37.27$ 
   & $> 3.6\times 10^{47}$& 17000 & 2.2$\times 10^{60}$
   & 1.2 / 11 \\
106$^{\dagger}$
   &              &          &$> 100^{a}$     
   & $< 2.7$      & $< 0.19$ & $> 37.10$ 
   & $> 2.4\times 10^{47}$& 19000 & 1.7$\times 10^{60}$
   & 1.5 / 11 \\ 
253$^{\dagger}$
   &              &          &$25\pm 2$
   & $14.5\pm 1.5$&$1.0\pm 0.1$&$36.14\pm 0.13$ 
   & $6.3 \sim 10^{46}$&$27000$ & $\lesssim 10^{59}$
   & 0.9 / 10 \\ 
614$^{\dagger,\,b}$
   &              &              & 17.2$^{c}$
   & $130\pm 15$  & $9.4\pm 0.9^{d}$ & $36.16\pm 0.13^{e}$
   &              &              &          
   &  3.6 / 13 \\ 
%
\hline
\end{tabular}
\end{center}
\vspace*{1mm}
$^{\star}$~according to the timing of the X-ray observations, 
$^{\dagger}$~according to the timing of the \textsl{IUE} 
             observations, 
$^{a}$~$= T_{\rm h}^{\rm min}$ (Sect.~3.2), 
$^{b}$~accretion disk (AD) model, 
$^{c}$~$= T^{\rm max}_{\rm disk}$ of the AD-model, 
$^{d}$~i.e. $R_{\rm disk} > 10$\ro, 
$^{e}$~$= L({\rm AD})/\cos(i)$.
\normalsize
\end{table*}
%
%
%
\begin{figure*}[t]
\centering
\begin{center}
\resizebox{15cm}{!}{\includegraphics[angle=-90]{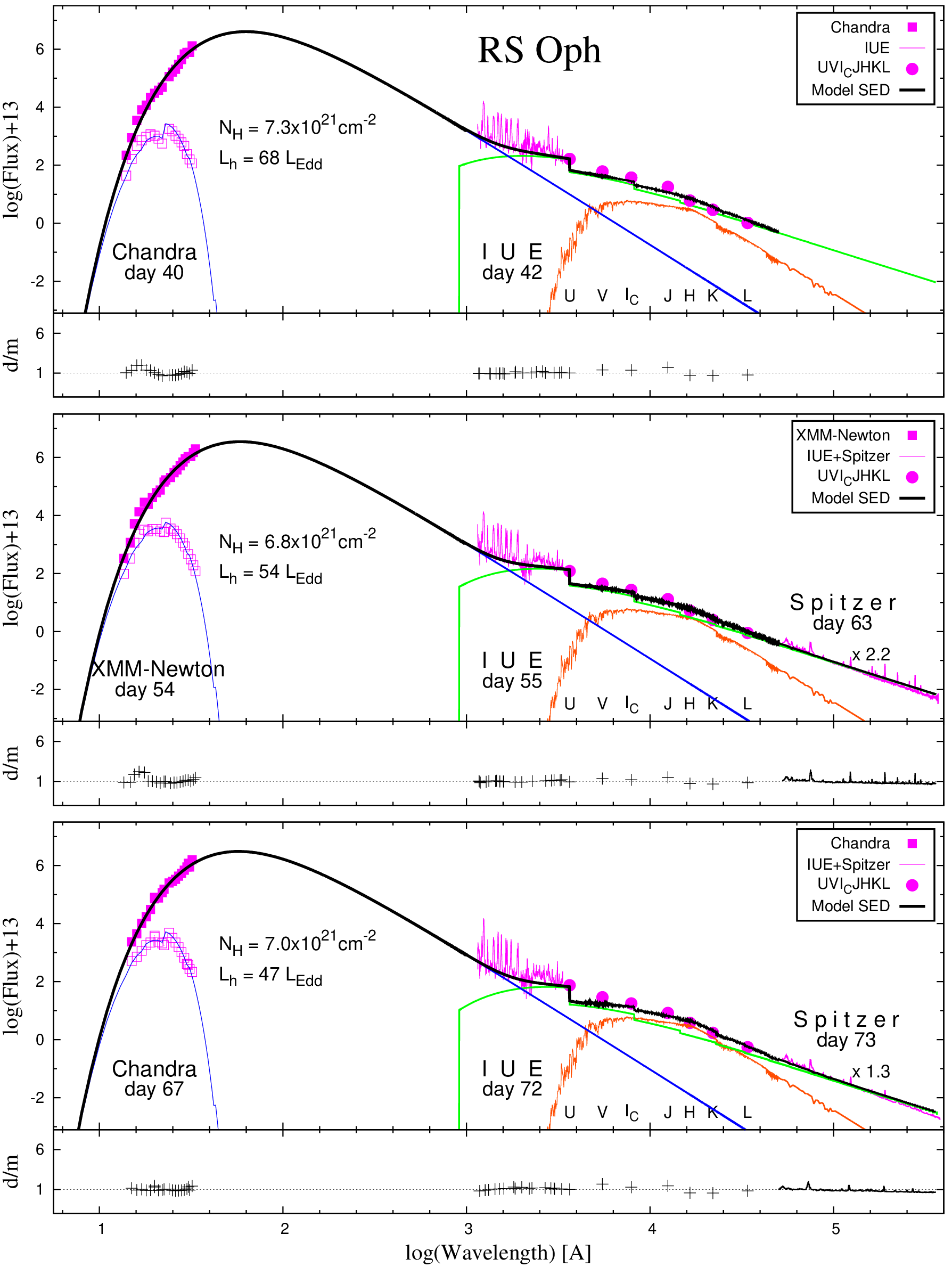}}
%
\end{center}
\caption[]{
A comparison of the observed (in violet) and model (heavy black 
line) SED of RS~Oph during its supersoft source phase with 
corresponding data-to-model ratios (d/m). Open/filled squares 
are the absorbed/unabsorbed X-ray fluxes. Fluxes are in units 
of \ecsa. The blue, green and orange line denotes the component 
of radiation from the WD, nebula and giant, respectively. 
          }
\label{fig:seds}
\end{figure*}

\section{Results of the SED-fitting analysis}

\subsection{Modelling the SED during the SSS phase}

I performed the the SED-fitting analysis in the same way as 
described in the section 2.6 of Paper~I. 
In addition, the observed grating spectra show a variable, but 
pronounced, jump absorption at $\sim 23.5$\,\AA, which is due 
to the oxygen absorption edge. According to \cite{ness+07}, 
changes in the \oi\ absorption edge profile at different days 
are caused by a different abundance of \oi\ atoms within the 
material in the line of sight. 
Therefore, to minimize the $\chi^{2}$ sum, I reduced the optical 
depth in the line of sight by a fractional contribution from 
oxygen, 
i.e. using $\tau_{\rm x} = \sigma_{\rm x} N_{\rm H} = 
(\sigma_{\rm ISM} - \xi \sigma_{\rm O\,I})N_{\rm H}$ in 
Eq.~(3) of Paper~I, 
where $\sigma_{\rm O\,I}$ is the \oi\ cross section in the ISM 
model reduced with a factor $0 < \xi < 1$. This factor thus 
represents an additional fitting parameter. 
Then, assuming that the interstellar part of the optical depth 
($N_{\rm H}(ISM) \sigma_{\rm ISM}$) persists as a constant, 
one can derive the abundance of the oxygen content within the 
{\em circumstellar} material in the line of sight as 
%
%
\begin{equation}
 A_{\rm O\,I}(CSM) = A_{\rm O\,I}(ISM)
         \left(1-\xi \frac{N_{\rm H}}{N_{\rm H}(CSM)}\right), 
\label{eq:aoi}
\end{equation}
where the oxygen abundance in the ISM, 
$A_{\rm O\,I}(ISM) = 0.00049$ \citep[][]{wilms+00}, 
and the hydrogen column density within the circumstellar 
matter, $N_{\rm H}(CSM) = N_{\rm H} - N_{\rm H}(ISM)$, 
as given by Eq.~(4) of Paper~I. 

For day 40 I fitted 18 X-ray fluxes from 14 to 32\,\AA, 
19 UV fluxes between 1170 and 3300\,\AA\ and the photometric 
\textsl{UV$I_C$JHKL} flux-points. I omitted flux-points 
derived from the \textsl{B} and \textsl{$R_C$} 
photometric measurements, because of a significant effect 
of emission lines in these passbands \citep[see][]{sk07}. 
The best-fit-model and the flux-point errors yielded the 
reduced $\chi^2_{\rm red}$ = 5.9 for 38 degrees of freedom. 
The larger value of $\chi^2_{\rm red}$ results mainly from 
small and thus uncertain fluxes from the short-wavelength 
part of the spectrum with the fixed 10\% errors only. 
In the same way I reconstructed the composite X-ray---IR 
continuum at days 54-55 and 67-73. All solutions required 
the total $N_{\rm H} > N_{\rm H}({\rm ISM})$ that reflects 
a significant contribution to {\rm b-f} absorptions by 
the CSM during the SSS phase of RS~Oph (Table~2). 

On day 40, the best-fit-model of the SSS 
suggested a reduction of the \oi\ content in 
the line of sight by $\sim 25$\% ($\xi\sim 0.25$), 
i.e. by $\sim 37$\% 
($= 0.25 N_{\rm H}/N_{\rm H}(CSM) \times 100$\%)
within the circumstellar matter. 
On day 54, the best fit suggested the parameter $\xi\sim 0.6$, 
which reduces oxygen within the CSM by more than 90\%. 
However, on day 67 the oxygen abundance $A_{\rm O\,I}(CSM)$ 
increased to $\sim 0.4 A_{\rm O\,I}(ISM)$. 
Similar results were found also by \cite{ness+07}. 

%
The \textsl{Spitzer} spectra could not be used in the fitting 
procedure, because they were not taken simultaneously with 
other modelled fluxes. Nevertheless, by multiplying them with 
an appropriate constant, the slope of their continuum followed 
that of the nebular component of radiation, which dominates 
the mid-IR domain (see Fig.~\ref{fig:seds}). 
Thus the SED-fitting analysis proved independently the nebular 
origin of the \textsl{Spitzer} spectra as suggested 
by \cite{evans07}. 

I made a rough estimate of the uncertainties in $T_{\rm h}$ and 
$N_{\rm H}$ by comparing more models around the best solution. 
All models with different $T_{\rm h}$ 
and $N_{\rm H}$ were scaled to the dereddened far-UV fluxes. 
In this way I estimated 
  $\Delta T_{\rm h}\sim 10000-20000$\,K 
and 
  $\Delta N_{\rm H}\sim 0.2-0.3\times 10^{21}$\,\cmd. 
The uncertainty for $\theta_{\rm h}$ was then derived with 
the aid of Eq.~(6) of Paper~I as the mean error of the 
$\theta_{\rm h}(N_{\rm H},T_{\rm h})$ total differential. 
Similarly, I derived the uncertainty in $L_{\rm h}$ as 
a function of $R_{\rm h}$ and $T_{\rm h}$ in the 
Stefan-Boltzman law. 
%
%
\begin{figure}[!t]
\centering
\begin{center}
\resizebox{\hsize}{!}{\includegraphics[angle=-90]{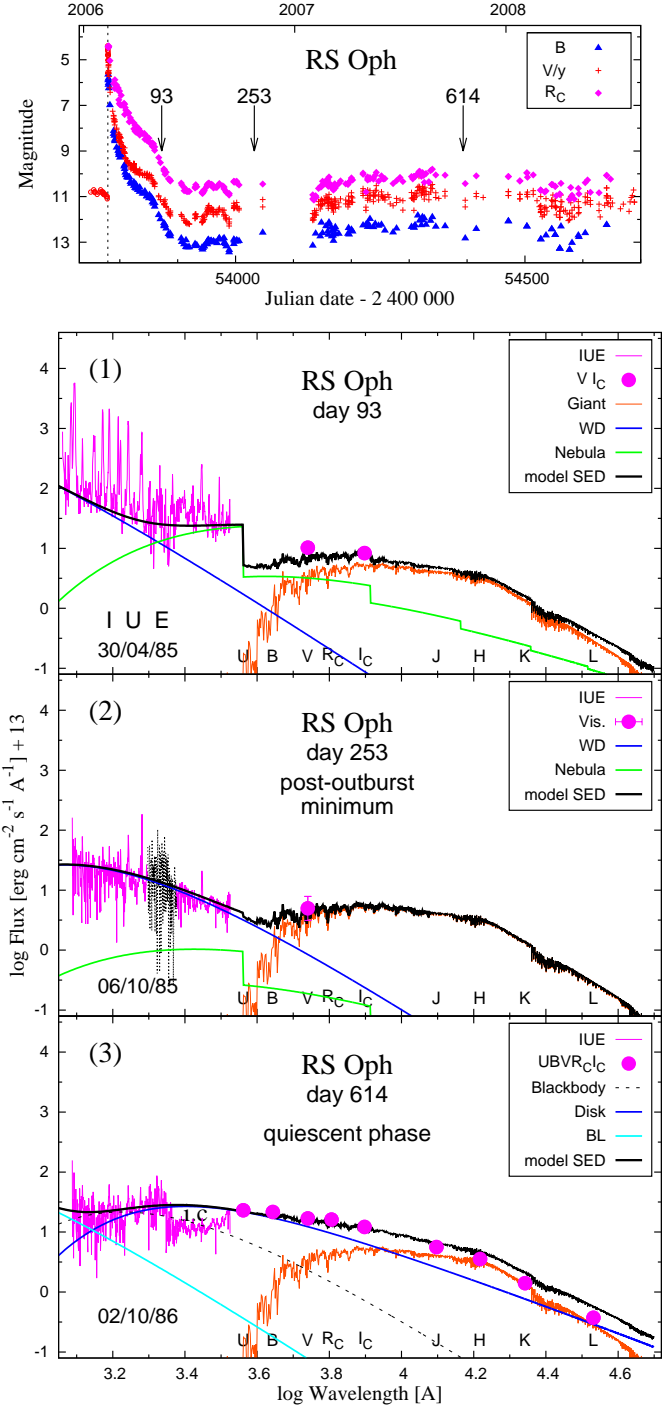}}
\end{center}
\caption[]{
Model SEDs beyond the SSS phase. (1) During the decline following 
the plateau phase (day 93), (2) during the post-outburst minimum 
(day 253) and (3) during the quiescence (day 614). Their timing 
is denoted by arrows in the top panel with the LC. Denotation of 
lines as in Fig.~\ref{fig:seds}. The dashed line in the bottom panel 
shows a Planck curve for a comparison (see Sect.~3.2). 
A depression of the UV continuum due to the iron curtain 
absorption is denoted by `i.c'. 
          } 
\label{fig:beyond}
\end{figure}  

\subsection{Modelling the SED beyond the SSS phase}

The model SED during the monotonic decline following the plateau 
phase (day 93.6 and 106.1) showed a significant decrease of both 
the stellar and the nebular component of radiation. 
Having only the UV part of the total WD spectrum, it was 
possible to determine only the minimum temperature, 
$T_{\rm h}^{\rm min} = 0.9-1\times 10^{5}$\,K, at which 
the WD's radiation gives rise to the observed \textsl{EM} and 
fits the far-UV fluxes (see Appendix~A of Paper~II). 
The \textsl{EM} of the nebula decreased by a factor of $\sim 5$ 
with respect to the day 67, but still dominated the optical. 

During the post-outburst minimum (day 253) the WD became 
significantly cooler, radiating at $\sim 25\,000$\,K 
and having an angular radius 
$\theta_{\rm h} \sim 1.45\times 10^{-11}$ 
(i.e. $R_{\rm h} \sim 1.03(d/1.6\,\kpc)$\ro\ and
      $L_{\rm h} \sim 370(d/1.6\,\kpc)^2$\lo).
The nebular emission was negligible with respect to 
contributions from the giant and WD (Fig.~\ref{fig:beyond}). 

During the quiescent phase (day 614) the SED profile throughout 
the UV/optical was more or less flat with the optical fluxes 
well above the radiation from the giant. A simple blackbody 
radiation could in no way match the flat SED, particularly not 
the optical (see Fig.~\ref{fig:beyond}, dashed line). 
Accordingly, I performed its modelling with a function, 
\begin{equation}
  F_{\rm h}^{\rm obs}(\lambda) = 
              \theta_{\rm h}^2(F_{\lambda}({\rm AD}) +
                 0.6 \pi B_{\lambda}(T_{\rm BL})), 
\label{eq:ad}
\end{equation}
where the first term at the right represents the flux 
distribution of an optically thick accretion disk that 
radiates locally like a black body, while the second term 
is a contribution from the boundary layer radiating 
at a temperature $T_{\rm BL}$. 
The disk temperature $T_{\star} = 2\,T_{\rm disk}^{\rm max}$ 
determines the slope of the UV continuum 
\citep[see e.g.][ in detail]{w95}. 
The observed flat UV continuum corresponds to 
  $T_{\rm disk}^{\rm max} = 17200$\,K, 
  $T_{\rm BL} \sim 79000$\,K and 
  $\theta_{\rm h} = 1.3\times 10^{-10}$. 
These parameters yield an effective (spherical) radius
  $R_{\rm h}^{\rm eff} \sim 9.4(d/1.6\,\kpc)$\ro, 
suggesting that the linear radius of the disk 
  $R_{\rm disk} > 10$\ro. The luminosity of the disk from 
the model SED is 
  $L_{\rm disk} \sim 380/\cos(i)$\lo\ and its boundary layer 
  $L_{\rm BL} \sim 700$\lo. However, the boundary layer 
is not well defined by the observed SED, because the far-UV 
fluxes are very uncertain. For example, their steepness could 
be simulated by an iron curtain absorption at longer wavelengths. 

The model SEDs are depicted in Figs.~\ref{fig:seds} and 
\ref{fig:beyond} and corresponding physical parameters 
are found in Table~2. 
%
%
\begin{figure}[!t]
\centering
\begin{center}
%
\resizebox{8cm}{!}{\includegraphics[angle=-90]{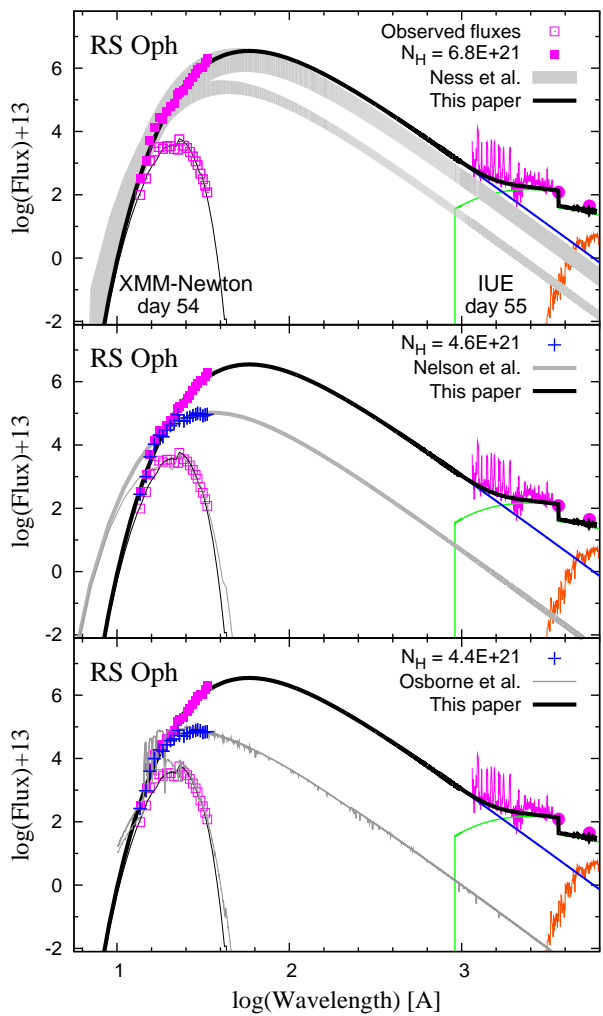}}
\end{center}
\caption[]{
Comparison of the multiwavelength model SED from day 54 
(Fig.~\ref{fig:seds}) and the X-ray-data models of 
\cite{ness+07} (top), \cite{nelson+08} (middle) and 
\cite{osborne+11} (bottom). 
          }
\label{fig:comparison}
\end{figure}

\section{Interpretation of observations}

The model SEDs throughout the 14\,\AA\---37\,$\mu$m range 
identified a strong {\em stellar} and {\em nebular} component 
of radiation in the spectrum of RS~Oph. 
The former is produced by the WD photosphere and dominates the 
spectrum from the X-rays to the far-UV, whereas the latter 
represents its fraction reprocessed by a thermal nebula via the 
b--f/(f--b and f--f) transitions under conditions of Case B. 
The nebula dominantes the spectrum for $\lambda \gtrsim 2000$\,\AA\ 
(Fig.~\ref{fig:seds}). 
However, the corresponding physical parameters are very different 
from those inferred from modelling only the X-ray data. 

\subsection{Comparison with previous models}

\cite{ness+07} first analyzed the grating spectra of RS~Oph 
obtained by \textsl{Chandra} and \textsl{XMM-Newton} during the SSS 
phase (Sect.~2). They developed a series of blackbody models to 
match the continuum emission, and found reasonable agreement with 
the measured spectra. The top panel of Fig.~\ref{fig:comparison} 
compares their models with that of this paper for day 54. The 
authors found the temperature and luminosity to be in the range 
of 
$T_{\rm h} = (650-710)\times 10^3$\,K and 
$\log(L_{\rm h}) = 38.5-38.9$, respectively, for 
  $N_{\rm H} \sim 5.3\times 10^{21}$\cmd\ (the lower shadow 
belt in the figure). By varying the oxygen abundance in the 
CSM component, they obtained a better agreement with 
  $T_{\rm h} = (520-580)\times 10^3$\,K, 
  $\log(L_{\rm h}) = 39.3-40.0$ 
and 
  $N_{\rm H} \sim 6.9\times 10^{21}$\cmd\ 
($N_{\rm H}$ according to Ness, private communication). The 
latter series of their models (see the upper shadow area in 
Fig.~\ref{fig:comparison}) is in a good agreement with the 
model of this paper. 
However, the authors did not claim from these models that 
a super-Eddington luminosity occurred. Because of modelling 
only the X-ray data, they were not able to decide on the 
reliability of their second sets of solutions. 

\cite{nelson+08} instead compared the observed spectra with 
atmospheric models calculated by \cite{rauch}. 
As the best-fit model they selected that with 
$T_{\rm h} = 820000 \pm 10000$\,K and 
$N_{\rm H} = 2.3\times 10^{21}$\cmd\ 
for the spectrum from day 54. Even fitting a blackbody to 
the spectra, they still obtained a temperature close to 800000\,K. 
Also these authors assumed that the WD luminosity does not 
exceed the Eddington value. The middle panel of 
Fig.~\ref{fig:comparison} shows that their blackbody model 
($T_{\rm h} = 800000$\,K, $L_{\rm h} \equiv L_{\rm Edd}$) 
is a factor of $\sim 150$ below the observed far-UV spectrum. 
In spite of this inconsistence, the X-ray data can still match 
their model with $N_{\rm H} \sim 4.6\times 10^{21}$\cmd\ 
(Fig.~\ref{fig:comparison}). 

\cite{osborne+11} performed spectral fits to the {\em Swift}-XRT 
spectra by both the blackbody and the atmosphere models. Although 
their plateau blackbody and model atmosphere luminosities were 
the same, they preferred the parameters given by the atmosphere 
models. During the plateau phase, around a maximum ($\sim$day 
54, see their Fig.~4), they obtained 
 $kT \sim 90$\,eV ($\sim 1.04\times 10^{6}$\,K), 
 $R \sim 4\times 10^8$\,cm ($\sim 0.0057$\ro), 
 $L \sim 3\times 10^4$\lo. 
They did not specify the corresponding $N_{\rm H}$. 
To determine its value, they used a power-law decrease of its 
CSM component \citep[according to][]{bode+06}, which should 
result in the total $N_{\rm H} = 3.4\times 10^{21}$\cmd\ on 
day 54. The authors also pointed out the gratifying agreement 
between their luminosities and those predicted theoretically 
by \cite{ibentutu96} for the plateau phase in the nova evolution. 
The bottom panel of Fig.~\ref{fig:comparison} shows an example 
of their atmospheric model 
(spectrum {\small {1020000-9.00-HHeCNONeMgSiS}} 
made on 2010-09-09 and 2011-01-26, available at 
\footnote{http://astro.uni-tuebingen.de/$^{\sim}$rauch/VO/fluxtables/ \\
HHeCNONeMgSiS\_gen/}), 
calculated for 
$T_{\rm eff} = 1.02\times 10^{6}$\,K and scaled to 
$L_{\rm h} = 2.5\times 10^{4}$\lo\ in the figure. However, the 
\textsl{XMM-Newton} fluxes had to be de-absorbed with 
$N_{\rm H} \sim 4.4\times 10^{21}$\cmd\ to match the model. 
Also in this case, the X-ray-data fit is far below the UV fluxes. 

In terms of the multiwavelength approach, the large differences 
between the parameters obtained by different groups of authors 
result from the well known mutual dependence between the 
parameters $L_{\rm h}$, $N_{\rm H}$ and $T_{\rm h}$ in fitting 
only the X-ray data: 
a larger/lower $L_{\rm h}$ requires a larger/lower $N_{\rm H}$ 
and a lower/higher $T_{\rm h}$ to fit the absorbed X-ray fluxes 
(see Sect.~4.1 of Paper~I). 

\subsection{Are there other sources contributing to the far-UV ?}

I consider the following two possibilities. 
(i) 
An accretion process at the super-Eddington rate of 
$\sim 3.6\times 10^{-5}$\myr\ (see below, Sect.~4.7.3) 
can release in maximum its binding energy of 
$\sim 1.4\times 10^{39}$\es\ ($M_{\rm WD} \sim 1.3$\mo, 
$R_{\rm WD} \sim 0.004$\ro; Sect.~4.6). The corresponding 
disk's luminosity is about twice the Eddington value. 
Its far-UV fluxes are well below the observed fluxes. 
(ii) 
The inverse Compton scattering of low energy photons off 
relativistic electrons cannot take place, because the plasma 
surrounding accreting WDs in the symbiotic novae and classical 
symbiotic stars contains non-relativistic electrons, as is 
clear from the presence of Thomson scattering measured in 
the strongest emission lines 
\citep[e.g. O\,VI 1032 and 1038\,\AA\ doublet, see][]{se+sk12}. 

If there is no other source, which contributes to the far-UV 
and rivals that from the burning WD, the very high far-UV fluxes 
($\sim 10^{-10}$\ecsa, see Fig.~\ref{fig:seds}) require the 
luminosity of the WD to be significantly above the Eddington 
limit. This can be probed independently by studying the very 
high emission from the thermal nebula. 
%
%
\begin{figure}[!t]
\centering
\begin{center}
%
\resizebox{\hsize}{!}{\includegraphics[angle=-90]{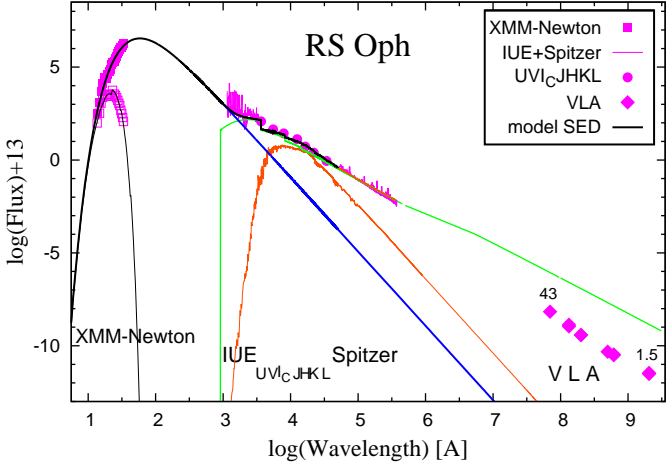}}
\end{center}
\caption[]{
The X-ray/radio SED for RS~Oph around day 55. Denotation of 
lines is as in Fig.~\ref{fig:seds}. Radio fluxes at 43, 22.5, 
15, 6, 5 and 1.5\,GHz were taken from \cite{sok+08} and 
\cite{eyres}. Their position shows that the nebula, which 
produces the UV/IR emission, is optically thick at radio 
frequencies (Sect.~4.3.1). The ${\rm f-f}$ Gaunt factor at 
radio frequencies was calculated according to approximation 
in \cite{kwok00}. 
          }
\label{fig:xradio}
\end{figure}

\subsection{Relations between the stellar and nebular emission}

The nebular emission represents a fraction of the stellar 
radiation from the WD reprocessed via the ionization/recombination 
and bremsstrahlung events. Therefore, its source can be associated 
with the material surrounding the burning WD. 

The quantity of the nebular emission was determined by the 
model SED and its extent was revealed by direct imaging 
\citep[][]{tob+06,rupen+08,sok+08}. At day 55, i.e. simultaneously 
with our second set of observations (Table~1), \cite{sok+08} 
observed RS~Oph with the Very Large Array (VLA) at 43\,GHz. 
They showed that the image emission is consistent with thermal 
bremsstrahlung emission from a lower temperature plasma heated 
by photoionizations. I used the image and the corresponding 
flux to estimate the volume of the emitting material and its 
average particle density, and test whether these parameters 
are consistent with the high WD's luminosity suggested by 
the model SED. 

\subsubsection{The emitting mass of the nebula}

Here I assume that the emitting material that is imaged at 
43\,GHz produces the nebular radiation we measure at UV--IR 
wavelengths. 
First, approximating the shape of the VLA image by a prolate 
spheroid with an axis 
$a \sim 34/{\rm sin}i$~mas = 71$(d/1.6\kpc)$\,AU 
and $b \sim 17.5$~mas = 28$(d/1.6\kpc)$\,AU, 
one obtains the volume of the main emission region, 
  $V^{neb} = 4/3\pi a b^2 \sim 
     7.8 \times 10^{44}(d/1.6\kpc)^3$\,cm$^{3}$. 
The emission measure observed at day 55, 
\textsl{EM}$ = 2.2\times 10^{61}(d/1.6\kpc)^2$\cmt, then yields 
an average particle density of this region, 
$\bar n \sim 1.7\times 10^{8}(d/1.6\kpc)^{-1/2}$\cmt. 
These quantities correspond to an emitting mass of 
the nebula 
  $M^{neb} = \mu m_{\rm H}\,\bar{n}\,V^{neb}$ 
$\sim 1.6\times 10^{-4}(d/1.6\kpc)^{5/2}$\mo, where $\mu \sim 1.4$ 
is the mean molecular weight and $m_{\rm H}$ is the mass
of the hydrogen atom. 

Second, using the measured flux at 43\,GHz, 
$F_{43} = 110 \pm 11$\,mJy \citep[][]{sok+08}, it is possible 
to verify results obtained from the image extension. Assuming 
that the plasma is optically thick at 43\,GHz, and radiates 
under conditions of thermodynamical equilibrium, the nebula 
resembles a blackbody at temperature $T_{\rm e}$. 
Then, using the Rayleigh-Jeans approximation 
($h\nu \ll kT_{\rm e}$, and no reddening applied), 
the angular radius of the nebula can be expressed as 
%
%
\begin{equation}
 \theta_{43} = \frac{c}{\nu}
               \left(\frac{\epsilon F_{43}}
                          {2 \pi k T_{\rm e}}\right)^{1/2} ,
\label{eq:th43}
\end{equation}
where the filling factor $\epsilon$ reduces the measured $F_{43}$ 
flux to a value produced only by the optically thick part of 
the nebula at 43\,GHz. Thus, $\epsilon = 1$ yields a maximum 
effective radius of the nebula, 
$R^{\rm eff}_{43} = (7.2\pm 0.4)\times 10^{14}(d/1.6\kpc)$\,cm, 
which corresponds to a spherical volume, $V^{\rm eff}_{43} = 
(1.6\pm 0.2)\times 10^{45}(d/1.6\kpc)^3$\,cm$^{3}$. 
For $\epsilon = 0.64$ both volumes are equal, 
$V^{\rm eff}_{43} = V^{neb}$. Their comparability verifies 
the above adopted assumptions. The optical depth for such 
an extended and dense nebula is as large as a few hundred 
at 43\,GHz \citep[see][ in detail]{kwok00}. The optically thick 
case for the RS~Oph remnant in the radio is demonstrated directly 
by the X-ray--radio SED (see Fig.~\ref{fig:xradio}). 
The radio fluxes are significantly lower than the nebular  
component of the radiation, which reproduces the UV--IR emission. 
So, if the measured \textsl{EM} corresponds to 
the $V^{\rm eff}_{43}$ volume, then 
$\bar n = (1.2\pm 0.1)\times 10^{8}(d/1.6\kpc)^{-1/2}$\cmt\ 
and 
$M^{neb}_{\rm 43} = 
  (2.1\pm 0.4)\times 10^{-4}(d/1.6\kpc)^{5/2}$\mo. 

Third, amount of the emitting mass can be proved independently 
by the hydrogen recombination lines, which are produced by the 
same volume of the nebula as the continuum. For example, using 
the \ha-line flux method \citep[e.g.][]{gurzadyan, kwok00}, 
the observed flux in the \ha\ line, $F_{\alpha}$, is 
produced by the ionized mass 
%
%
\begin{equation}
 M_{\alpha}^{neb} = 4\pi d^2 \frac{\mu m_{\rm H}}
              {\varepsilon_{\alpha}(T_{\rm e}) n_{\rm e}}\,
                   F_{\alpha} ,
\label{eq:ma-neb}
\end{equation}
where $\varepsilon_{\alpha}(T_{\rm e})$ is the volume emission 
coefficient in the \ha\ line. 
\cite{sk+08} analyzed the profile of a strong \ha\ line taken at 
day 57 (see their Fig.~3). According to Eq.~(\ref{eq:ma-neb}), 
its integrated flux, $F_{\alpha} = (6.5 \pm 0.1)\times 10^{-9}$\ecs, 
implies an emitting mass 
$M_{\alpha}^{neb} = 1.1-1.6\times 10^{-4}(d/1.6\kpc)^{5/2}$\mo\ 
for 
$n_{\rm e} \equiv \bar n = 1.7-1.2\times 10^{8}(d/1.6\kpc)^{-1/2}$\cmt, 
as derived above from the nebular continuum. The volume emission 
coefficient $\varepsilon_{\alpha}(30000) = 1.24\times 10^{-25}$ 
${\rm erg\,cm^{3}\,s^{-1}}$ \cite[e.g.][]{ost89}. 
The range of $M_{\alpha}^{neb}$ values represents rather a lower 
limit, because the densest central part of the nebula, which 
contributes mainly to the line core, can be optically thick. 

I conclude that: (i) the large quantity of \textsl{EM} derived 
from the SED, (ii) the volume and the flux of the emitting region 
measured by the VLA, and (iii) the strong flux in the \ha\ line 
from the optical spectrum, all correspond to a high mass of 
the nebular emitting material, 
$M^{neb} = (1.6\pm 0.5)\times 10^{-4}(d/1.6\kpc)^{5/2}$\mo, 
which surrounds the burning WD even during the SSS phase of 
its outbursts. 

\subsubsection{Constraints for super-Eddington luminosity}

(i) 
The nebular emission can be created only within the ionized 
part of the material surrounding the ionizing source. Its extent 
is limited by the distance from the WD's photosphere, at which 
the flux of its ionizing photons is balanced by the rate of 
ionization/recombination events inside the nebula. The radius 
of such an ionized zone is known as the Str\"omgren 
sphere, $r_{\rm S}$. Here I express it in the form 
\citep[see Eq.~(3) of][]{sk+09} 
%
%
\begin{equation}
 r_{\rm S} = \left(\frac{3 L_{\rm ph}({\rm H})}
             {4\pi \alpha_{\rm B}({\rm H},T_{\rm e})}\,
             \bar{n}^{-2} \right)^{1/3},
\label{eq:rs}
\end{equation}
where $L_{\rm ph}({\rm H})$ is the flux of photons capable 
of ionizing hydrogen, 
$\alpha_{\rm B}({\rm H},T_{\rm e})$ [cm$^{3}$\,s$^{-1}$] 
stands for the total hydrogenic recombination coefficient in 
Case $B$, and $\bar{n}$ is the mean particle density. 
The Eddington luminosity of the burning WD (44000\lo\ for 
$M_{\rm WD} = 1.3$\mo) radiating at $8-10\times 10^{5}$\,K 
\citep[][]{nelson+08,osborne+11} corresponds to 
$L_{\rm ph}({\rm H}) = 5.6-4.5\times 10^{47}$\,s$^{-1}$. 
Then $\bar n = 1.7\times 10^{8}$\cmt\ (Sect.~4.3.1) and 
$\alpha_{\rm B} = 1\times 10^{-13}$\,${\rm cm^{3} s^{-1}}$ 
yield $r_{\rm S} = 24-22$\,AU, which is, however, smaller 
than the observed extent of the nebular material 
($a = 71$ and $b = 28$\,AU, Sect.~4.3.1). Thus, the 
WD radiating at the Eddington luminosity is not capable 
to ionize the observed volume of the material emitting 
at 43\,GHz, whose image was used to determine $\bar n$. 
This implies that $L_{\rm h} > L_{\rm Edd}$ to increase 
$L_{\rm ph}({\rm H})$ and thus $r_{\rm S}$. 

(ii) 
A lower limit of the WD luminosity can be estimated from 
the nebular component of radiation, which represents a fraction 
of the ionizing source radiation reprocessed by the thermal 
nebula via the ionization/recombination events. 
If {\em all} the ionizing photons are converted into the nebular 
radiation, then the {\em lower} limit of the WD luminosity, which 
is capable of producing the observed \textsl{EM}, can be 
expressed as (see Appendix~A of Paper~II), 
%
%
\begin{figure}[t]
\centering
\begin{center}
%
\resizebox{\hsize}{!}{\includegraphics[angle=-90]{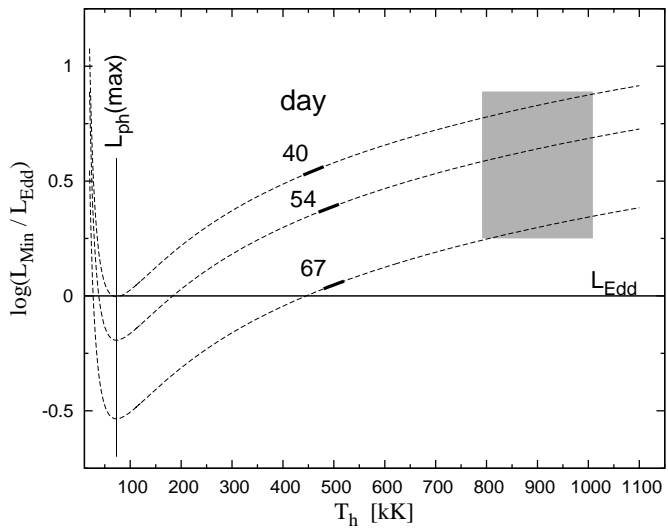}}
\end{center}
\caption[]{
The lower limit of the WD luminosity (Eq.~(\ref{eq:lmin})) as 
a function of its temperature scaled with \textsl{EM} as given 
by the model SED on day 40, 54 and 67. Solid thick bars bound 
ranges of $T_{\rm h}$ (Table~2). 
The gray belt denotes temperatures derived from the 
X-ray-data models \citep[][]{nelson+08,osborne+11}. 
Vertical line points the minimum of the $L_{\rm Min}(T_{\rm h})$ 
function at $T_{\rm h} = 73000$\,K, for which the source produces 
a maximum flux of hydrogen ionizing photons, 
$L_{\rm ph}({\rm max})$ (see Sect.~4.3.2). 
          }
\label{fig:lmin}
\end{figure}
%
%
\begin{equation}
 L_{\rm Min}(T_{\rm h}) = \alpha_{\rm B}({\rm H},T_{\rm e})\,
                          \textsl{EM}
                          \frac{\sigma T_{\rm h}^{4}}{f(T_{\rm h})}, 
\label{eq:lmin}
\end{equation}
%
where the function
%
%
\begin{equation}
f(T_{\rm h}) = \frac{\pi}{hc}\int^{\rm 912\AA}_{0}\!\!\!
               \lambda\, B_{\lambda}(T_{\rm h})\,\rm d\lambda.
\label{eq:fth}
\end{equation}
Equation~(\ref{eq:lmin}) is valid for the hydrogen plasma, 
heated by photoionizations, and characterized with a constant 
$T_{\rm e}$ and $\bar{n}$. Thus, the \textsl{EM} constrains 
a minimum luminosity of the ionizing source as a function 
of its temperature. Figure~\ref{fig:lmin} shows the 
$L_{\rm Min}(T_{\rm h})$ function parameterized with the 
\textsl{EM} obtained from the model SED on day 40, 54 and 67. 
The minimum occurs at $T_{\rm h}\sim 73000$\,K, at which  
the source produces a maximum $L_{\rm ph}({\rm H})$. 
Higher values of $L_{\rm Min}(T_{\rm h})$ on both sides reflect 
lower values of $L_{\rm ph}({\rm H})$, which thus need a higher 
$L_{\rm h}(T_{\rm h})$ of the source to give rise 
the measured \textsl{EM}. 
The steep increase of $L_{\rm Min}$ towards low $T_{\rm h}$ 
is due to a rapid decrease in $L_{\rm ph}({\rm H})$. 
A gradual increase of $L_{\rm Min}$ towards high $T_{\rm h}$ 
reflects a gradual decrease in $L_{\rm ph}({\rm H})$, 
because a lower {\em number} of higher energy photons gives 
the same luminosity in \es. 

Figure~\ref{fig:lmin} thus demonstrates that the high quantity 
of the \textsl{EM} constrains $L_{\rm Min} > L_{\rm Edd}$ 
for the temperatures given by model SEDs. 
It also shows that the parameters of the SSS, as derived from 
the X-ray-data models ($T_{\rm h}\sim 8-10 \times 10^5$\,K, 
$L_{\rm h} \lesssim L_{\rm Edd}$, Sect.~4.1), are not consistent 
with the observed amount of nebular emission. The values of 
\textsl{EM} require 
$L_{\rm Min}(8-10 \times 10^5$\,K) = 2--7$L_{\rm Edd}$ 
(see the intersection of the gray array with the dashed lines in 
Fig.~\ref{fig:lmin}). In other words, $L_{\rm h}$ and $T_{\rm h}$ 
obtained from the X-ray-data fits are not capable of producing 
the observed nebular emission. The \textsl{EM} of the thermal 
plasma thus represents a critical parameter in estimating the 
lower limit of the SSS luminosity. 

In addition, a fraction of the ionizing photons can escape the 
star without being converted into the nebular radiation, and also 
the measured \textsl{EM} represents a lower limit of what was 
originally created by ionizations, because we observe only 
the optically thin part of the nebula.
Under these circumstances the flux of ionizing photons, originally 
emitted by the star, is larger than that given by the equilibrium 
condition. This implies 
\begin{equation}
   L_{\rm h} > L_{\rm Min} > L_{\rm Edd}.
\label{eq:lhgt}
\end{equation}
(see Appendix~A of Paper~II in detail). 
In this way the large value of the \textsl{EM} justifies 
independently the WD's super-Eddington luminosity derived 
from the multiwavelength model SEDs, even during the SSS phase. 

\subsection{The nature of the plateau phase}

The plateau phase in the LC developed approximately between 
day 40 and 80 after the optical maximum, and is coincident 
with the SSS phase \citep[][]{hkl07}. Figure~\ref{fig:seds} 
shows that the optical $UBVRI$ region is dominated by 
the nebular continuum during this phase. 
A relatively slow decrease of the optical light thus reflects 
a relatively slow decrease in the \textsl{EM}. 
The variation in the \textsl{EM} follows solely that in the 
luminosity, because it is a function of the rate of 
$L_{\rm ph}$ photons. 
As a result, the slow decrease in the luminosity during the SSS 
phase causes a relevant decrease in the \textsl{EM}, which causes 
only a slow fading in the optical brightness, indicated in the LC  
as the {\em plateau phase} (see Fig.~1). 
Therefore, the cessation of the hydrogen burning around day 80, 
which stops the strong SSS phase \citep[][]{hkl07}, ends also 
the plateau phase in the LC, because of a significant 
reduction of the WD's luminosity, i.e. production of the 
$L_{\rm ph}$ photons. Model SEDs beyond the SSS phase confirm 
this conclusion (Table~2, Fig.~\ref{fig:beyond}, Sect.~4.6). 

Finally, I note that the {\em nebular} nature of the optical 
continuum during the SSS phase contradicts modelling the plateau 
phase in the LC with the {\em blackbody} radiation 
produced by a large disk irradiated by the WD photosphere 
as proposed by \cite{hachisu+06}. 

\subsection{The growing mass of the WD due to its outbursts}

The WD luminosity and the mass liberated during the burning 
phase ($\sim 80$ days) allow us to estimate the amount of 
hydrogen burnt, and thus the mass added to the WD surface 
during the outburst, $M_{\rm add}^{burnt}$. 
The prime energy source is the fusion of four protons that 
produces one helium atom. The mass difference between the input 
fuel and the output ash reveals that 0.7\%\ of the mass of 
the original protons is converted into energy, 
released in the form of gamma rays and neutrinos. This means 
that one gram of hydrogen can generate energy of 
$\eta = 0.007\times c^2 \sim 
            6.3 \times 10^{18}$\,erg\,g$^{-1}$ 
\citep[details of the hydrogen burning efficiency can be found 
in][]{mit89}. 
To estimate $M_{\rm add}^{burnt}$ I will assume that the energy 
generated by the thermonuclear fusion balances just the observed 
luminosity and the energy required to lift off the mass, 
$\Delta M_{\rm wind}$, expelled via the wind, i.e. 
\begin{equation}
M_{\rm add}^{burnt} = 0.993\,\mu 
      \left(\frac{L_{\rm h} \Delta t}{\eta}  + 
      \frac{G M_{\rm WD}\Delta M_{\rm wind}}
           {\eta \,R_{\rm WD}}\right) ,
\label{eq:madded}
\end{equation}
where $\mu \sim 1.4$ is the mean molecular weight and 
$\Delta t = 80$\,days. 

Modelling the broad \ha\ wings from the beginning of the eruption 
(day 1.38) and from the SSS phase (day 57) \citep[see][]{sk+08}, 
and the evolution in the line profiles along the outburst 
\cite[e.g.][]{iijima,banerjee}, suggested the mass 
loss rate via the wind at $\approx 10^{-4}$\myr\ for the first 
10 days, and $\approx 10^{-5}$\myr\ during the following 
70 days until the end of the hydrogen burning phase. Thus, 
the mass ejected from the accumulated WD envelope can be 
roughly estimated to 
$\Delta M_{\rm wind}\approx 10/365\times 10^{-4} + 70/365 
\times 10^{-5} = 4.6\times 10^{-6}$\mo. 
This value is a factor of $\sim$1.6 larger than that estimated 
by \cite{hkl07}. However, they assumed that the wind from the 
WD stopped already at the beginning of the SSS phase. 

Assuming constant $L_{\rm h} \sim 1\times 10^{40}$\es\ (Table~2), 
using parameters of a high-mass WD and $\Delta t$ and 
$\Delta M_{\rm wind}$ as above, Eq.~(\ref{eq:madded}) yields 
$M_{\rm add}^{burnt} \sim 8\times 10^{-6}$\mo\ for $d = 1.6$\kpc. 
For the outburst's recurrence time of 20 years, the WD mass 
in RS~Oph is thus growing at an average rate of 
$\dot M_{\rm WD} \sim 4\times 10^{-7}(d/1.6\kpc)^2$\myr. 
However, a certain fraction of the WD material can be 
dredged up and mixed into the ejecta due to the TNR. 
This effect can decrease the $M_{\rm add}^{burnt}$ mass 
and thus also its growing rate. 
Modelling the supersoft X-ray and visual LCs, \cite{hkl07} 
derived $\dot M_{\rm WD} \sim 1\times 10^{-7}$\myr. 

\subsection{Evolution beyond the SSS phase}

Following the plateau phase in the LC (day$\gtrsim 80$), both 
the stellar radiation from the WD and the nebular radiation 
were decreasing (Table~2, Sect.~3.2). A relatively strong 
contribution from the nebula (in both the continuum and lines), 
as measured around the middle of the monotonic decline, 
(day 93, Fig.~\ref{fig:beyond}) implied that the WD still 
emitted a large flux of ionizing photons at the rate 
$\gtrsim 3\times 10^{47}$\,s$^{-1}$. 

During the post-outburst minimum (day 253, Fig.~\ref{fig:beyond}), 
the WD photosphere became markedly cooler 
($T_{\rm h}\sim 25000$\,K) and larger 
($R_{\rm h}^{\rm eff}\sim 1$\ro). The nebular contribution 
was negligible. As a result, the corresponding SED showed 
a minimum at the optical (see the figure). 
This effect was transient, because the following cooling of 
the hot component (i.e. the WD photosphere and the disk-like 
material surrounding it) caused filling-in the optical during 
the quiescent phase ($\gtrsim 1$\,year after the outburst). 

During the quiescence, the flat UV/optical SED satisfied 
radiation produced by a large ($R_{\rm disk} > 10$\ro) 
optically thick accretion disk (day 614, Fig.~\ref{fig:beyond}). 
In addition, the pronounced features of the iron curtain in 
the \textsl{IUE} spectra constrain the presence of the veiling 
material located at/behind the outer rim of the disk, and 
extended vertically from its plane, so to cause the observed 
absorptions. This naturally explains a significant depression 
of the X-ray luminosity indicated during the quiescent 
phase \citep[][and references therein]{osborne+11}. 
These properties imply that the disk-like formation in RS~Oph 
is not geometrically thin as accretion disks in CVs. 
However, the satisfactory model SED with Eq.~(\ref{eq:ad}) 
suggests that the warm pseudophotosphere in RS~Oph is also 
given by an ansamble of contributions radiating at different 
temperatures (probably along its projection to the sky). 

The luminosity of the disk ((280 -- 510)/$\cos(i)$\lo, Table~2) 
corresponds to an accretion rate of $\dot M_{\rm acc}(Q) = 
(1.6 - 3.0)\times 10^{-7}$\myr\ for $\alpha = 0.5$ 
\citep[][]{starf+88}, $M_{\rm WD} \sim 1.3$\mo, 
$R_{\rm WD} \sim 0.004$\ro\ and $i = 50^{\circ}$ 
\citep[][]{nauenberg,yaron,brandi+09}. 
This implies that the high mass WD in RS~Oph is accreting 
just below the stable burning limit 
\citep[e.g.][]{shen+bild07,wolf+13}. 
During the 20 years of quiescence the WD accumulates the mass 
of $m_{\rm acc}(Q) = (3.2 - 6.0)\times 10^{-6}$\mo, which 
is sufficient to ignite a new outburst 
(see Eq.~(\ref{eq:pbase})). 

Finally, according to the above-mentioned disk properties, 
the parameters $L_{\rm h}$ and $\dot M_{\rm acc}(Q)$ can be 
a factor of $1/\cos(i)$ lower, but also can be somewhat larger, 
because a fraction of the radiaton from the inner parts of 
the disk in the direction of poles cannot be detected 
\citep[see Sect.~5.3.6 of][]{sk05}. 
However, values of $L_{\rm h}$ and $T_{\rm h}$ from our model 
SED are consitent with those estimated by \cite{dobrzycka+96}, 
who classified the hot component in RS~Oph during quiescence 
as a B-type shell star. 

\subsection{On the accretion and ejection in RS~Oph}

Recently, \cite{schaefer09} discussed in detail the accretion 
process in RS~Oph, and concluded that wind accretion fails 
to provide enough mass to the WD required for an eruption. 
Therefore he assumed that 
the giant in RS~Oph fills its Roche lobe, and placed the 
system to the distance of $4.3\pm 0.7$\kpc. However, this 
assumption is not consistent with the model SED of the red 
giant in RS~Oph. 

The large luminosity of the giant for the distance of 
4.3\kpc, $L_{\rm g} = 690\,(4.3/1.6\kpc)^2 \sim 5000$\lo\ 
(see Sect.~3.1 of Paper~II), 
is not compatible with its small mass ${\cal M}_{\rm g}$ 
= 0.68--0.8\mo\ \citep[][]{brandi+09} and the effective 
temperature $T_{\rm eff} = 3800-4200$\,K (Paper~II). 
Note that both ${\cal M}_{\rm g}$ and $T_{\rm eff}$ are 
distance independent. 
According to the current calibration of (super)giants 
\citep[see][]{cox} and/or evolutionary models for single stars 
\citep[e.g.][]{loore}, such stellar parameters are not 
consistent with any normal star. 

Accordingly, the giant in RS~Oph cannot fill its Roche lobe, 
and thus the required high accretion rate cannot be a result 
of a standard Roche lobe overflow (see Sect.~4.7.3). 

\subsubsection{On the origin of the massive nebula}

The huge amount of the emitting mass, 
$M^{neb} \gtrsim 10^{-4}(d/1.6\kpc)^{5/2}$\mo, which surrounds 
the binary, persists in the system until the end 
of the hydrogen burning phase (Sect.~4.3.1). 
The presence of a significant amount of CSM in the system 
is also consistent with the high value of $N_{\rm H}$ during 
the whole SSS phase, which is by a factor of $\sim$3 larger 
than the interstellar value. 
%
Does this emitting CSM in RS~Oph come from the accreted WD envelope? 
The answer is {\em no}. If this were the case, then the pressure 
at the base of the envelope, 
\begin{equation}
 P_{\rm base} = G \frac{M_{\rm WD} M_{\rm env}}
                       {4\pi R_{\rm WD}^4}, 
\label{eq:pbase}
\end{equation}
would be $\sim 10^{21}\,{\rm dyne\,cm^{-2}}$ for parameters of 
a high-mass WD, which exceeds significantly the critical value, 
$P_{\rm crit} \approx 10^{19}$\,dyne\,cm$^{-2}$, 
required to ignite a TNR \citep[e.g.][]{yaron}. 
This would also imply an unrealistically high average accretion 
rate ($\dot M_{\rm acc} \sim M_{\rm env}/P_{\rm rec} 
\sim 10^{-5}$\myr). 
Nevertheless, RS~Oph ejected a fraction of the accreted mass, as 
was proved by extremely broad emission lines, observed at 
the beginning of its outburst \citep[][]{buil,tarasova,iijima}. 
In such a case the nuclear energy, $E_{\rm nuc}$, has to be 
larger than the gravitational potential energy of the central WD, 
to lift a mass $m_{\rm ej}$ outside the system. 
This leads to a condition 
\begin{equation}
  m_{\rm ej} < \frac{R_{\rm WD}}{G M_{\rm WD}} E_{\rm nuc} ,
\label{eq:mej}
\end{equation}
where $E_{\rm nuc} = L_{\rm h} \Delta t$ and $m_{\rm ej}$ 
is the mass released during the time $\Delta t$. 
Parameters, $L_{\rm h} \sim 10^{40}$erg\,s$^{-1}$, 
$\Delta t = 54$\,days, $M_{\rm WD} \sim 1.3$\mo\ 
and $R_{\rm WD} \sim 0.004$\ro\ (see above) 
yield 
       $m_{\rm ej} < 4 \times 10^{-5}$\mo. 
So, the emitting mass of $\gtrsim 10^{-4}(d/1.6\kpc)^{5/2}$\mo, 
as measured on day 54, could not be ejected from the WD surface. 

Therefore, the large amount of mass, emitting during the RS~Oph 
outbursts, has to have its origin outside the accreted envelope, 
most probably in a large disk encompassing the WD. 

\subsubsection{Evidence for a disk around the WD}

(i) 
During first $\sim$30 days after the explosion, a bipolar 
jet-like collimated outflow was indicated by the satellite 
components to the main cores of hydrogen line profiles 
\citep[][]{sk+08,banerjee}. Later, directly seen in the 
43\,GHz image \citep[][]{sok+08}. This indicates the presence 
of a disk, because jets require an accretion disk 
\citep[e.g.][]{livio97}. 

(ii)
Photometric flickering variability became to be measurable 
shortly after the outburst, at day 117 \citep[][]{zamanov+06} 
and later on, increasing in the amplitude, through the whole 
period between outbursts \citep[][]{dobrzycka+96,worters,hric+08}. 
It is believed that flickering originates from accretion onto 
a WD, and thus its source can be located within the accretion 
disk \citep[e.g.][]{bruch92}. 

(iii) 
A pronounced double-peak structure in the \ha\ profile was 
observed between nova eruptions \citep[][]{zamanov+05,brandi+09}. 
After the 2006 outburst, a signature of this type of the profile 
was first measured on day 200 (2006, August 31) by \cite{iijima}. 
The origin of the double-peak profile of hydrogen Balmer lines 
observed in symbiotic stars is often connected with 
an accretion disk \citep[e.g.][]{robinson}. 

(iv)
The flux distribution from the UV to near-IR, during quiescent 
phase, could be modelled directly by the radiation from a large 
optically thick accretion disk ($R_{\rm disk} > 10$\ro, 
Sects.~3.2 and 4.6, Fig.~\ref{fig:beyond}). 
Assuming a typical average density in the disk 
$\log(\rho) \approx -7$ in g\cmt, the mass of the disk can 
be as high as $\approx 10^{-4}$\mo. 
It is of interest to note that the SED of RS~Oph from quiescence 
is very similar to that of classical symbiotic stars during 
active phases, which requires a disk-like formation around 
the WD \citep[e.g. Figs.~4, 5, 6, 27 of][]{sk05}. 

\subsubsection{On the accretion mechanism}

The mass added on the WD surface due to hydrogen burning and 
the mass ejected during the outburst represent a minimum to 
the material accreted by the WD, 
$m_{\rm acc} \sim M_{\rm add}^{burnt} + \Delta M_{\rm wind} 
             = 1.26\times 10^{-5}$\mo\ (Sect.~4.5). 
The high quantity of $m_{\rm acc}$ 
constrains the average accretion rate to 
$\dot M_{\rm acc} \equiv m_{\rm acc}/P_{\rm rec} 
\sim 6.3\times 10^{-7}$\myr\ for $P_{\rm rec} = 20$\,years. 
During quiescence, the matter from the disk is accreted at 
$\sim 2.3\times 10^{-7}$\myr\ (Sect.~4.6), 
which implies enhanced accretion during the 80-day burning 
phase to $\sim 3.6\times 10^{-5}$\myr\ to balance the average 
accretion between outbursts. 
Such a high accretion rate during the outburst could be 
realized through the disk, as indicated by the presence of 
jets (Sect.~4.7.2). 

The final problem in reconstructing the picture of the RS~Oph 
outburst is to identify the source of material, which feeds 
the WD at the required rate. Comparison of the giant's 
luminosity in RS~Oph (Sect.~3.1 of Paper~II) with those 
in other symbiotic stars, suggests a wind mass loss rate from 
the giant, $\dot M_{\rm giant, wind} \sim 10^{-7}$\myr\ 
\citep[see Fig.~24 in][]{sk05}. If the efficiency of the 
accretion process from the wind is (in maximum) of 10\%\ 
\citep[][]{nagae}, then the required 
$\dot M_{\rm acc} \gg 0.1\times \dot M_{\rm giant, wind}$, 
which shows that the giant itself is not capable of supplying 
the necessary material to the WD in a standard wind accretion 
regime \citep[see][in detail]{schaefer09}. 
In connection with this problem, I will consider two 
possibilities. 

(i) 
The only vital source of material for the accretion is the large 
amount of emitting CSM indicated during the hydrogen burning 
phase (Sect.~4.3.1). The relatively fast re-creation of the 
large disk in the system (see Sect.~4.6, Fig.~\ref{fig:beyond}) 
suggests that this material has to return back to the binary, 
where it enhances the accretion onto the WD. Observationally, 
this process is indicated directly by the A-type absorption 
lines, whose radial velocities indicate a movement of absorbing 
material towards the binary at a velocity of $\gtrsim 5$\kms\ 
\citep[see Fig.~3 and Table~2 of][]{brandi+09}. 

In this case, when the accretion from the giant's wind is 
neglected, the large accretion rate from the disk-like reservoir 
of material implies a gradual decrease of the disk's mass and 
thus accretion. 
As a result, the recurrence time of the outbursts will extend, 
the growing of the WD mass will decelerate, and thus prolong 
the time, when the WD mass will reach the Chandrasekhar limit. 
The extending period (9, 18 and 21 years) between the most 
recent 4 explosions (1958, 1967, 1985 and 2006) supports 
this case. Accordingly, the next outburst should occur 
beyond 2027. 

(ii)
The primary source of the material powering the outburst is 
the wind from the giant. In this case the wind has to be 
compressed to the equatorial plane to fill in continuously 
the disk, and thus to keep the accretion rate at the required 
level. 
As a result, outbursts of RS~Oph should be rather periodic, 
lasting for a long time, until the WD reach its limiting mass. 
In this case, the next explosion can occur even prior to 2027. 

A possibility of a very effective mass transfer mode in binaries 
containing an evolved star -- wind Roche-lobe overflow -- was 
recently suggested by \cite{moh+pod07}. This mode can be 
in the effect, when the acceleration zone of the wind extends 
to the Roche lobe \citep[see Fig.~1 of][]{abate+13}. Then, the 
wind of the giant is focused towards the orbital plane and in 
particular towards the WD, which then accretes at a significantly 
larger rate than from the spherically symmetric wind 
\citep[][]{moh+pod07,borro+09}. A relatively small Roche-lobe 
radius of the giant in RS~Oph, 
$R_{\rm L,g} \sim 2\times R_{\rm g}$ (see Paper~II), is in 
favour of this case. 
Another possibility how to compress the wind from a star to 
the equatorial plane is the rotation of the central star 
\citep[][]{bjorkcass93}. 
\cite{zam+stoy12} found that giants in symbiotic stars rotate 
with a mean $v\sin(i)\sim 8$\kms. Using this value and other 
characteristic parameters of the giant's wind in symbiotic 
stars (e.g. $\dot M_{\rm giant, wind} \sim 10^{-7}$\myr\ as 
above and terminal velocity of 20--40\kms), the wind compression 
at the orbital plane and a distance of 2--3 A.U. from the wind 
source yields $\dot M_{\rm giant, wind} \gtrsim 10^{-6}$\myr\ 
(Carikov\'a and Skopal, in preparation). Such a high mass-loss 
rate should be capable of feeding the disk with 
a few times $10^{-7}$\myr. 

\section{Summary}

In this paper I investigated the 14\,\AA\ to 37\,$\mu$m continuum 
radiation emitted by the recurrent symbiotic nova RS~Oph during 
its SSS phase and the following transition to quiescence. 
For this purpose I used the method of multiwavelength modelling 
of the SED (see Paper~I). Particularly, I disentangled 
the composite continuum at the beginning (day 40), middle 
(day 54) and the end (day 67) of the stable SSS phase, and during 
the final decline (day 93), the post-outburst minimum (day 253) 
and the quiescence (day 614). 
During the SSS phase, the model SEDs revealed the presence of 
a very strong stellar as well as nebular component of radiation 
in the spectrum. The former was emitted by the burning WD, while 
the latter represents a fraction of its radiation reprocessed by 
the thermal nebula. 
During the transition to quiescence, both components were 
decreasing and during quiescence the SED satisfied radiation 
produced by a large ($R_{\rm disk} > 10$\ro) optically thick 
accretion disk. 
The model SEDs are depicted in Figs.~\ref{fig:seds} and   
\ref{fig:beyond} and the corresponding physical parameters
are found in Table~2. 
The main results of this paper can be summarized as follows. 
\begin{enumerate}
\item
{\em The stellar} component of radiation can be reproduced 
by that of a blackbody with an average effective radius 
  $R_{\rm h}^{\rm eff}\sim 0.23 (d/1.6\kpc)$\ro, 
radiating at the temperature
  $T_{\rm h}\sim 490\,000$\,K. 
This yields a luminosity 
  $L_{\rm h}\sim 10^{40} (d/1.6\kpc)^2$\es, 
which exceeds the Eddington luminosity by a factor of 
$\sim 60$. 
The X-ray fluxes were attenuated by the {\rm b-f} absorptions 
corresponding to the total neutral hydrogen column density
  $N_{\rm H} \sim 7\times 10^{21}$\cmd, which exceeds 
the interstellar value by a factor of $\sim$3. 
This reflects a significant additional attenuation by 
the CSM (Sect.~3.1, Fig.~\ref{fig:seds}, Table~2). 
%
{\em The nebular} component of radiation dominated 
the spectrum from the mid-UV to longer wavelengths. 
It was characterized with an electron temperature 
  $T_{\rm e} \sim 30\,000$\,K 
and a large emission measure, 
  \textsl{EM}$ \sim 2.2\times 10^{61}(d/1.6\kpc)^2$\cmt. 
It was present in the spectrum throughout the whole SSS phase 
(Fig.~\ref{fig:seds}, Table~2). 
{\em The radiation from the giant} was described in Paper~II. 
Here, an additional argument that the giant underfills 
its Roche lobe is found in Sect.~4.7. 
\item 
The parameters of the SSS, as determined by the multiwavelength 
modelling the global SED, are very different from those obtained 
by fitting only the X-ray data. However, the X-ray-data models 
do not match the UV fluxes at all 
(Sect.~4.1, Fig.~\ref{fig:comparison}). 
Such a difference is given by the well known mutual dependence 
between the parameters $L_{\rm h}$, $N_{\rm H}$ and $T_{\rm h}$, 
which allows to obtain a wide scale of solutions in fitting only 
the X-ray data (see Sect.~4.1 of Paper~I). Therefore, the 
authors selected that satisfying the current theoretical 
predictions (e.g. $L_{\rm h} \lesssim L_{\rm Edd}$ during the SSS 
phase). 
\item
The super-Eddington luminosity during the SSS phase is 
independently supported by the high quantity of the nebular 
emission, which represents a fraction of the WD radiation 
reprocessed by the thermal nebula 
(Sect.~4.3.2, Fig.~\ref{fig:lmin}, Eq.~(\ref{eq:lhgt})). 
\item 
The plateau phase in the LC is caused by a slow fading of the 
nebular radiation that dominates the optical. As the nebula 
was fed by the WD radiation, the cessation of the hydrogen 
burning around day 80, which stops the SSS phase, ends also 
the plateau phase in the LC (Sect.~4.4). 
\item
The large value of the \textsl{EM}, the volume and the flux of 
the emitting region as well as the strong flux in the \ha\ line 
correspond to a high mass of the nebular emitting material, 
$M^{neb} = (1.6\pm 0.5)\times 10^{-4}(d/1.6\kpc)^{5/2}$\mo\ 
(Sect.~4.3.1). 
It has its origin outside the accreted envelope, in a large 
disk encompassing the WD (Sect.~4.7.1). 
\item
The WD's luminosity and the energy required to lift off the 
mass from the WD during the burning phase correspond to the 
mass added on the WD surface due to hydrogen burning, 
$M_{\rm add}^{burnt} \sim 8\times 10^{-6}(d/1.6\kpc)^2$\mo. 
For the outburst's recurrence time of 20 years, the WD mass 
in RS~Oph is thus growing at an average rate of 
$\dot M_{\rm WD} \sim 4\times 10^{-7}(d/1.6\kpc)^2$\myr\ 
(Sect.~4.5). 
\item
The mass accreted by the WD between outbursts, 
$m_{\rm acc} \sim 1.26\times 10^{-5}$\mo, constrains the average 
accretion rate 
$\dot M_{\rm acc} \sim 6.3\times 10^{-7}$\myr\ (Sect.~4.7.3). 
\item
During quiescence, the model SED requires an average accretion 
from the disk, $\dot M_{\rm acc}(Q) \sim 2.3\times 10^{-7}$\myr, 
which implies that the WD in RS~Oph is accreting just below 
the stable burning limit. During the 20 years of quiescence, 
the WD thus accumulates the mass of 
$m_{\rm acc}(Q) \sim 4.6\times 10^{-6}$\mo, which is sufficient 
to ignite a new explosion (Sect.~4.6). 
\item
During the 80-day burning phase, the accretion from the disk 
has to be enhanced to $\sim 3.6\times 10^{-5}$\myr\ (Sect.~4.7.3). 
Such a high transient accretion rate is also signalized by the 
transient presence of jets (Sect.~4.7.2). It can be thus 
responsible for the super-Eddington luminosity during the 
whole hydrogen burning phase. 
\item
If the wind from the giant is {\em not} sufficient to feed 
the WD at the required rate, the accretion has to be realized 
from a disk-like reservoir of mass in the system. 
During the burning phase, the massive nebula can represent such 
the reservoir of the mass (Sects.~4.3.1 and 4.7.1). 
During the quiescent phase, the presence of a large and massive 
disk is indicated by the model SED (Sect.~4.7.2, point (iv)). 
In this case, the time between the outbursts will extend. The 
next outburst should occur beyond 2027, according to 
timing of the last 4 well observed explosions (1958, 1967, 
1985, 2006, $> 2027$; Sect.~4.7.3, point (i)). 
\item
If the primary source of the high accretion rate is the wind 
from the giant, then it has to be focused towards the orbital 
plane to fill in continuously the disk. Then, the next explosion 
can occur prior to 2027 (Sect.~4.7.3, point (ii)). 
\end{enumerate}
The solution of the multiwavelength modelling the global 
SED of the recurrent symbiotic nova RS~Oph suggests that 
the mass transfer and accretion via the wind from the 
evolved star onto its compact companion in a wide binary 
must be more effective than the currently accepted view. 

\section*{Acknowledgments}
I thank the anonymous referee for constructive comments. 
I am grateful to Horst Drechsel for a discussion 
to the early version of this work and all arrangements during 
my visits at the Astronomisches Institut der Universit\"at 
Erlangen-N\"urnberg in Bamberg (2008, 2010 and 2012). 
I thank Manfred Hanke for providing me the recent cross-sections 
for the X-ray absorption model and some advice about treatment 
the X-ray data. 
Nye Evans is thanked for providing me the \textsl{Spitzer} 
spectra of his original PID~270 proposal in a table form. 

This work is in part based on observations obtained with 
\textsl{XMM-Newton}, an ESA science mission with instruments 
and contributions directly funded by ESA Member States and 
NASA. 
This research has also made use of data obtained from the 
{\em Chandra} X-ray observatory operated for NASA by the 
Smithsonian Astrophysical Observatory. 
This work profited from enormous effort of other astronomers, 
who gathered a large amount of multivavelength observations 
throughout the very wide wavelength range. 

Finally, a partial support of this research was realized through 
a grant of Alexander von Humboldt foundation No.~SLA/1039115 and 
a grant of the the Slovak Academy of Sciences, VEGA No.~2/0002/13. 
%
%

%
\end{document}